\documentclass[prb, aps, floatfix, twocolumn, superscriptaddress]{revtex4-2}
\def\be{\begin{equation}}
\def\ee{\end{equation}}
\def\bea{\begin{eqnarray}}
\def\eea{\end{eqnarray}}
\def\bi{\begin{itemize}}
\def\ei{\end{itemize}}
\usepackage{braket}
\usepackage{xcolor}
\usepackage{graphicx}
\usepackage{amsmath,amstext,amssymb,bm,color,times}
\usepackage[normalem]{ulem} 
\usepackage[english]{babel}
\usepackage[T1]{fontenc}
\usepackage[utf8]{inputenc}
\usepackage[colorlinks=true,citecolor=blue,linkcolor=magenta]{hyperref}

\begin{document}


\title{
Kibble-Zurek dynamical scaling hypothesis \\
in the Google analog-digital quantum simulator of the $XX$ model
}

\newcommand{\affilju}{
             Jagiellonian University, 
             Faculty of Physics, Astronomy and Applied Computer Science,
             Institute of Theoretical Physics, 
             ul. \L{}ojasiewicza 11, 30-348 Krak\'ow, Poland 
             }
             
\newcommand{\affildoc}{ 
             Jagiellonian University, 
             Doctoral School of Exact and Natural Sciences, 
             ul. \L{}ojasiewicza 11, 30-348 Krak\'ow, Poland 
             }             

\newcommand{\affilkac}{  
             Jagiellonian University, 
             Mark Kac Center for Complex Systems Research,
             ul. \L{}ojasiewicza 11, 30-348 Krak\'ow, Poland 
             }
             
\author{Yintai Zhang}\affiliation{\affildoc}\affiliation{\affilju}
\author{Francis A. Bayocboc Jr.}\affiliation{\affilju}
\author{Jacek Dziarmaga}\affiliation{\affilju}\affiliation{\affilkac}

\date{\today}

\begin{abstract}
State-of-the-art tensor networks are employed to simulate the Hamiltonian ramp in the analog-digital quantum simulation of the quantum phase transition to the quasi-long-range ordered phase of the two-dimensional square-lattice $XX$ model [T.I. Andersen \textit{et al.}, Nature (London) \textbf{638}, 79 (2025)]. We focus on the quantum Kibble-Zurek (KZ) mechanism near the quantum critical point. 
Using the infinite projected entangled pair state, we simulate an infinite lattice and demonstrate the KZ scaling hypothesis for the $XX$ correlations across a wide range of ramp times.    
We use the time-dependent variational principle algorithm to simulate a finite $8\times 8$ lattice, similar to the one in the quantum simulation, and find that adiabatic finite-size effects dominate for longer ramp times, where the correlation length's growth with increasing ramp time saturates and the excitation energy's dependence on the ramp time crosses over to a power-law decay characteristic of adiabatic transitions. This finding contradicts the quantum simulation data where the correlation length seems to obey KZ-like power laws, although with modified exponents.
\end{abstract}

\maketitle


\section{Introduction}
\label{sec:intro}

The Kibble-Zurek mechanism (KZM) originated from a scenario for topological defect formation in cosmological phase transitions, where the independent selection of broken symmetry vacua in causally disconnected regions results in a mosaic of broken symmetry domains, leading to topologically nontrivial configurations~\cite{K-a, *K-b, *K-c}. 
For phase transitions in condensed matter systems, relativistic causality is not relevant, and a dynamical theory for continuous phase transitions was proposed~\cite{Z-a,*Z-b,*Z-c,Z-d}. It predicts the scaling of the defect density as a function of the quench rate by employing the universality class of the transitions. It has been verified by simulations~\cite{KZnum-a,KZnum-b,KZnum-c,KZnum-d,KZnum-e,KZnum-f,KZnum-g,KZnum-h,KZnum-i,KZnum-j,KZnum-k,KZnum-l,KZnum-m,that,KZ_weakly_first} and experiments~\cite{KZexp-a,KZexp-b,KZexp-c,KZexp-d,KZexp-e,KZexp-f,KZexp-g,QKZexp-a,KZexp-h,KZexp-i,KZexp-j,KZexp-k,KZexp-l,KZexp-m,KZexp-n,KZexp-o,KZexp-p,KZexp-q,lamporesi2013,donadello2016,KZexp-s,KZexp-t,KZexp-u,KZexp-v,KZexp-w,KZexp-x}. 
The quantum version of KZM (QKZM) was developed for quenches across quantum critical points in isolated systems~\cite{QKZ1,QKZ2,QKZ3,d2005,d2010-a, d2010-b,QKZteor-a,QKZteor-b,QKZteor-c,QKZteor-d,QKZteor-e,QKZteor-f,QKZteor-g,QKZteor-h,QKZteor-i,QKZteor-j,QKZteor-k,QKZteor-l,QKZteor-m,QKZteor-n,QKZteor-o,KZLR1,KZLR2,QKZteor-oo,delcampostatistics,KZLR3,QKZteor-q,QKZteor-r,QKZteor-s,QKZteor-t,sonic,QKZteor-u,QKZteor-v,QKZteor-w,QKZteor-x,roychowdhury2020dynamics,sonic, schmitt2021quantum,RadekNowak,dziarmaga_kinks_2022,transverse_oscillations,persistent_osc} and tested by experiments~\cite{QKZexp-a, QKZexp-b, QKZexp-c, QKZexp-d, QKZexp-e, QKZexp-f, QKZexp-g,deMarco2,Lukin18,adolfodwave,2dkzdwave,King_Dwave1d_2022,Semeghini2021,Satzinger2021etal}. Recent progress in Rydberg atoms' versatile emulation of quantum many-body systems~\cite{rydberg2d1,rydberg2d2,Semeghini2021,Satzinger2021etal} and coherent D-Wave~\cite{King_Dwave1d_2022,King_Dwave_glass,Science_Dwave} has opened the possibility to study the QKZM in a variety of two- and three-dimensional (2D and 3D) settings and/or to employ it as a test of quantumness of the hardware~\cite{RadekNowak, King_Dwave1d_2022, dziarmaga_kinks_2022, schmitt2021quantum,transverse_oscillations,persistent_osc,dwave_mitigation}.

\begin{figure}[t!]
    \includegraphics[width=0.99\columnwidth]{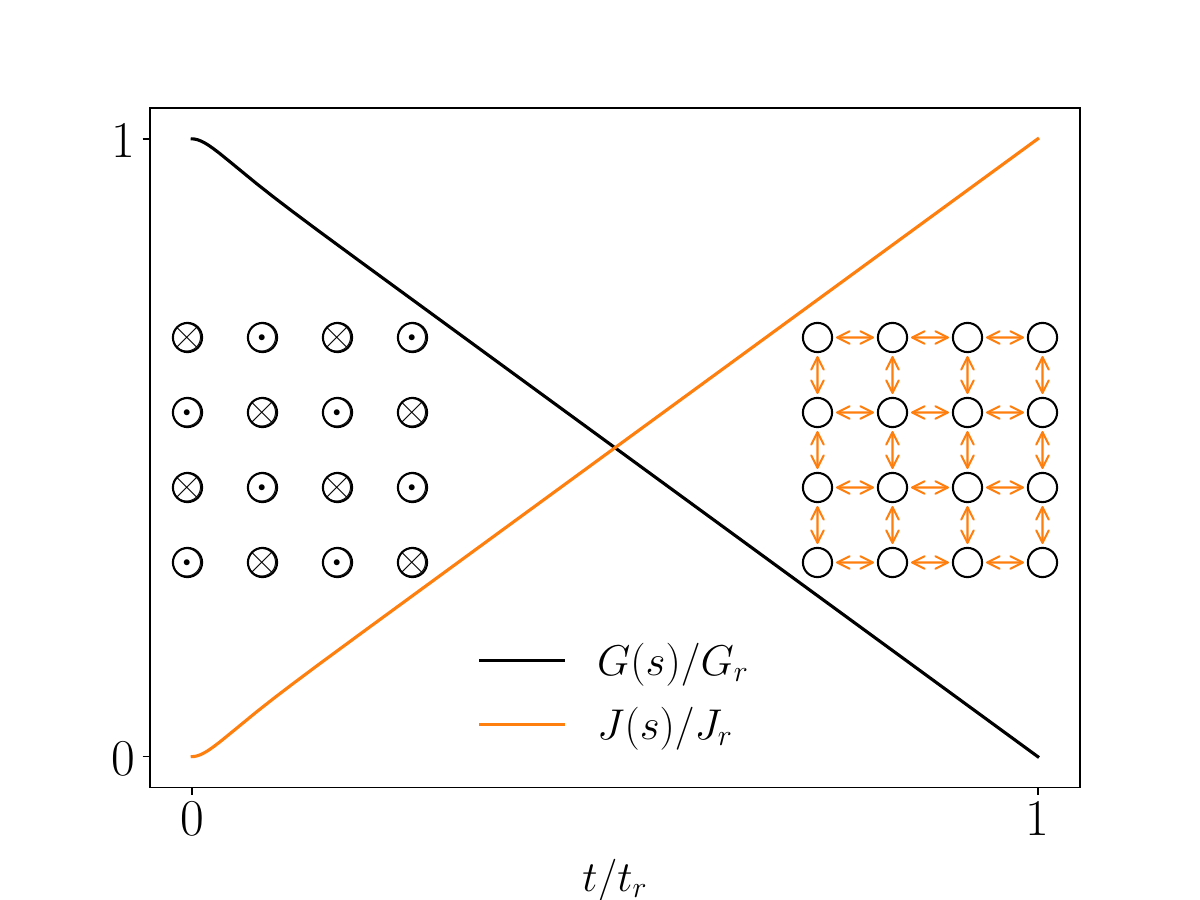}
    \caption{{\bf The ramp.}
    The values of $G(s)$ and $J(s)$ in the Hamiltonian \eqref{eq:Hsigma}.
    The experimental ramp~\cite{XX_Google} is linear $s = (t/t_r)$.
    In our tensor network simulation, it is slightly modified to $s = (t/t_r)[1-\exp(-40t/t_r)]$ to reduce excitation at the beginning of the ramp~\cite{Science_Dwave}. 
    The left inset shows the initial state of a small $4 \times 4$ system with a staggered field of strength $G(0)=G_{r}$ and nearest-neighbor (NN) coupling $J(0) = 0$. The dots (crosses) correspond to fields going out of (into) the page. The right inset shows the system at the end of the ramp where the staggered field is turned off, $G(1)=0$, and the nearest-neighbor coupling is turned on, $J(1) = J_r$. 
    }\label{fig:ramp}
\end{figure}

In a recent experiment~\cite{Lukin_coarsening}, a quench across a quantum critical point in the 2D transverse-field quantum Ising model, simulated with Rydberg atoms, was performed to study the slow quantum coarsening dynamics of domain walls following their excitation when crossing the critical point. This coarsening stage lies beyond the capabilities of current state-of-the-art classical numerical methods \cite{schmitt2021quantum}.
In a parallel experiment using a novel analog-digital quantum simulator~\cite{XX_Google}, a quantum phase transition from a paramagnetic phase to the quasi-ferromagnetic phase of the 2D quantum XX model was driven in order to study thermalization of the QKZ excitations above and below the Kosterlitz-Thouless critical temperature~\cite{XX_Google} (see Fig. \ref{fig:ramp}). This is an interesting attempt at a programmable quantum simulation on a universal quantum computer, an endeavor of high potential civilization impact. For the first time, digital and analog capabilities are combined in the same hardware. It is therefore of utmost importance to assess the quality of the simulation and see how far it goes beyond what is simulable on classical computers. In this paper, we use tensor networks to simulate QKZM near the critical point. 

QKZM can be briefly outlined as follows (for more thorough reviews, see Refs.~\onlinecite{d2010-a,d2010-b,Z-d,ROSSINI20211}).
A smooth ramp crossing the critical point at time $t_c$ can be linearized in its vicinity as 
\begin{equation}
\epsilon(t)=\frac{t-t_c}{\tau_Q}.
\label{epsilont}
\end{equation}
Here, $\epsilon$ is a dimensionless parameter in the Hamiltonian that measures the distance from the quantum critical point, and $\tau_Q$ is the quench time. Initially, the system is prepared in its ground state. Far from the critical point, the evolution adiabatically follows the ground state of the time-dependent Hamiltonian. However, adiabaticity must break down near the time $t_c-\hat t$, when the energy gap becomes comparable to the quench ramp rate: 
\begin{equation}
\Delta\propto|\epsilon|^{z\nu} \propto |\dot \epsilon/\epsilon| = 1/|t-t_c|. 
\end{equation}
This timescale is 
\be 
\hat t\propto \tau_Q^{z\nu/(1+z\nu)},
\label{hatt}
\ee
where $z$ and $\nu$ are the dynamical and correlation length critical exponents, respectively. 
The ground state correlation length at $t_c-\hat t$, 
\begin{equation}
\hat\xi \propto \tau_Q^{\nu/(1+z\nu)}, 
\label{hatxi}
\end{equation}
defines the typical size of the domains in which fluctuations select the same symmetry-broken ground state. 
The two Kibble-Zurek (KZ) scales are related by
\be   
\hat t \propto \hat\xi^z.
\ee 
Accordingly, in the KZM regime after $t_c-\hat t$, observables are expected to satisfy the KZM dynamical scaling hypothesis~\cite{KZscaling1,KZscaling2,Francuzetal}, with $\hat\xi$ being the unique scale. For a two-point observable ${\cal O}_R$, where $R$ is the distance between the two points, this reads
\be 
\hat\xi^{\Delta_{\cal O}} \bra{\psi(t)} {\cal O}_R \ket{\psi(t)} = 
F_{\cal O}\left[ (t-t_c)/\hat\xi^z , R/\hat\xi \right],
\label{eq:KZscalingO}
\ee
where $\ket{\psi(t)}$ is the state during the quench, $\Delta_{\cal O}$ is the scaling dimension of the operator ${\cal O}$, and $F_{\cal O}$ is a nonuniversal scaling function. 
Beyond the KZM ``cartoon picture,'' the correlation length does not remain frozen between $-\hat t$ and $+\hat t$, but continues to grow \cite{sonic}. When the dynamical exponent $z=1$, its growth rate is limited by the relevant speed of sound at the critical point. Further corrections to the leading KZM power laws were predicted in Ref.~\onlinecite{KZ_correction}.  

\begin{figure}[t!]
    \includegraphics[width=0.99\columnwidth]{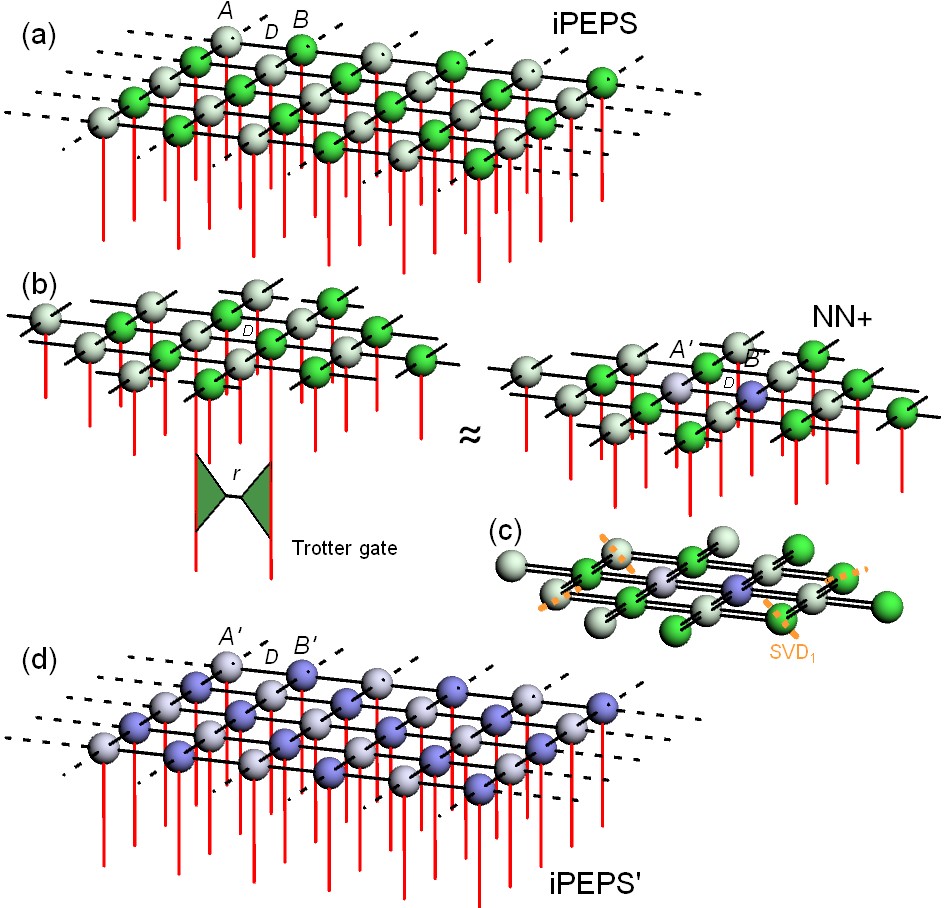}
    \caption{{\bf Trotter gate.}
    In (a), an infinite PEPS (iPEPS) tensor network with two sublattice tensors $A$ and $B$ and bond dimension $D$.
    In (b) left, a two-site Trotter gate is applied to a pair of tensors, increasing the bond dimension from $D$ to $rD$. 
    In (b) right, the dimension is truncated back to $D$. The error of the truncation is the Frobenius norm of the difference between the left and the right. The two tensors on the right, $A'$ and $B'$, are optimized to minimize the error.
    In (c), the norm squared of (b) right. Here, the balls are double PEPS tensors (contractions of PEPS tensors with their conjugates). The four corner doubles are approximated by their singular value decompositions truncated to one singular value (SVD$_1$). 
    In (d), the optimized $A'$ and $B'$ make the new iPEPS' ready for application of the next Trotter gate. 
    }\label{fig:nn+}
\end{figure}

\begin{figure}[t!]
    \includegraphics[width=0.97\columnwidth]{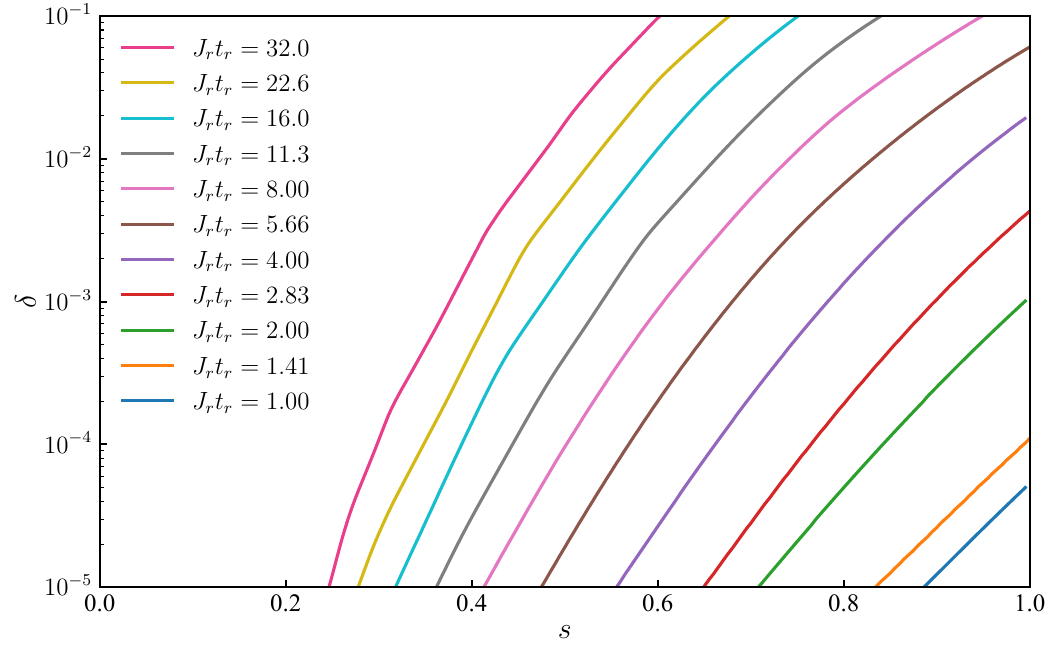}
    \caption{{\bf Accumulated truncation error.}
    The accumulated truncation error $\delta$ [see \eqref{eq:delta}] as a function of the ramp parameter $s$ for different ramp times and maximal bond dimension $D=24$.
    It is a rough upper estimate for errors of local observables inflicted by TN truncation~\cite{Science_Dwave}.
    }\label{fig:error}
\end{figure}

\section{Experiment and simulation}
\label{sec:exp_and_TN}

The experiment~\cite{XX_Google} can be described by the Hamiltonian 
\be
H =
\frac12 J \sum_{\langle i,j \rangle} \left( \sigma^x_i\sigma^x_j + \sigma^y_i\sigma^y_j \right) +
\frac12 G \sum_j h_j \sigma_j^z.
\label{eq:Hsigma}
\ee
Here, $h_j=\pm 1$ is a staggered transverse field. The antiferromagnetic coupling and the field strength are ramped as follows, respectively:
\bea 
J(s) = s J_r,~~ G(s) = \left( 1 - s \right)G_r
\eea
(see Fig. \ref{fig:ramp}). Here, $s=t/t_r$ is a linear ramp, $t_r$ is the ramp time, and $J_r=2\pi\times20$\,MHz and $G_r=2\pi\times30$\,MHz are the ramp magnitudes. The initial state at $t=0$ is the N\'{e}el ground state for $J=0$. The best numerical estimate of the critical point in Ref.~\onlinecite{XX_Google} is $G_c/J_c=1.8(6)$, which translates to $s_c\approx0.45(8)$. The universality class of this continuous quantum transition is expected to be that of the 3D XY universality class, because it is related to the spontaneous breaking of its $U(1)$ global symmetry. In the following, we use its critical exponents: $z=1$ and Monte Carlo estimates $\eta=0.03810(8)$ and $\nu=0.67169(7)$~\cite{Hasenbusch} (see also Tab. 3 in Ref.~\onlinecite{ROSSINI20211}, where other estimates can be found).

Here, we simulate the same ramp with a 2D infinite projected entangled pair states (iPEPS) tensor network (TN)~\cite{nishino01,gendiar03,verstraete2004,verstraete2004, Murg_finitePEPS_07,Cirac_iPEPS_08,Xiang_SU_08,Orus_CTM_09,fu,Lubasch_conditioning,Corboz_varopt_16, Vanderstraeten_varopt_16, Fishman_FPCTM_17, Xie_PEPScontr_17, Corboz_Eextrap_16, Corboz_FCLS_18, Rader_FCLS_18, Rams_xiD_18},
as shown in Fig. \ref{fig:nn+} (a). 
The iPEPS has been used to simulate sudden Hamiltonian quenches \cite{CzarnikDziarmagaCorboz,HubigCirac,tJholeHubig,SUlocalization,SUtimecrystal,ntu,mbl_ntu,BH2Dcorrelationspreading,ising2D_correlationsperading,Mazur_BH,Corboz_SF}. 
Given PEPS's noncanonical structure, it is necessary to resort to local updates in time evolution, such as the neighborhood tensor update (NTU) \cite{ntu}. NTU has been used to simulate many-body localization \cite{mbl_ntu}, Kibble-Zurek ramps in the quantum Ising and the Bose-Hubbard models \cite{schmitt2021quantum,Mazur_BH,Science_Dwave}, the bang-bang preparation of ground states by shallow quantum circuits \cite{BB_iPEPS,BB2}, as well as thermal states obtained by imaginary time evolution in the fermionic Hubbard model \cite{Hubbard_Sinha,Sinha_Wietek_Hubbard}. 
In this work, we use the second-order Suzuki-Trotter decomposition and NTU with the NN+ neighborhood (see Fig. \ref{fig:nn+}) introduced in Ref.~\onlinecite{Science_Dwave} in the context of testing the quantum computational advantage. Unlike in the spin glasses in Ref.~\onlinecite{Science_Dwave}, here the KZ excitations are not localized and they spread entanglement across the system, making the tensor network simulation more challenging to further test the new method. The $U(1)$ symmetry underlying conservation of $\sum_j\sigma_j^z$ is employed within the YASTN package~\cite{YASTN1}. We set the time step as $dt=\min(0.001\ \mu\mathrm{s}, 0.005t_r)$ and, in order to verify the KZ power laws, consider a geometric progression of ramp times with $J_rt_r$ ranging from $1$ to $32$. 

The expressive power of the iPEPS is limited by its bond dimension $D$. Here, we employ $D$ up to $24$ and use it as a refinement parameter to assess if observables have converged with increasing $D$. Furthermore, the accuracy is monitored using the truncation error, defined as the Frobenius norm of the difference between the left and the right diagram in Fig. \ref{fig:nn+} (b). The norm divided by the norm of the exact left diagram defines a relative truncation error $\delta_i$ of the $i$th Trotter gate applied to the bond. In the worst-case scenario of additive errors, the accuracy can be characterized by the accumulated truncation error \cite{Hubbard_Sinha}:
\be 
\delta=\sum_i\delta_i.
\label{eq:delta}
\ee 
It provides a rough upper estimate of relative errors for local observables. Figure \ref{fig:error} shows $\delta$ as a function of the ramp parameter $s$ for different ramp times. The longer ramp times are harder to simulate because they allow more time for any excitations to propagate and spread entanglement across the system. Their simulation has to be terminated at earlier $s$, but it can still cover the KZ regime near the critical point to verify the KZ scaling hypothesis \eqref{eq:KZscalingO}.

For a finite lattice, we use the time-dependent variational principle (TDVP) algorithm~\cite{Haegeman2011} in YASTN~\cite{YASTN1} and check convergence of observables by increasing the matrix product state's (MPS) bond dimension $D$ up to $1024$.

\begin{figure}[t!]
    \includegraphics[width=0.99\columnwidth]{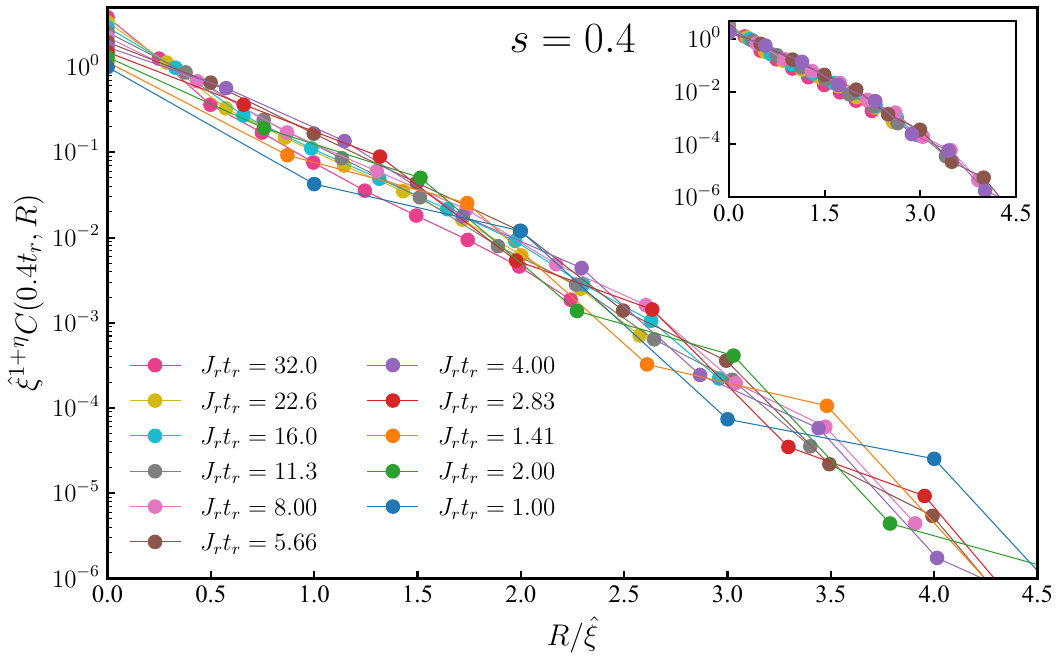}
    \includegraphics[width=0.99\columnwidth]{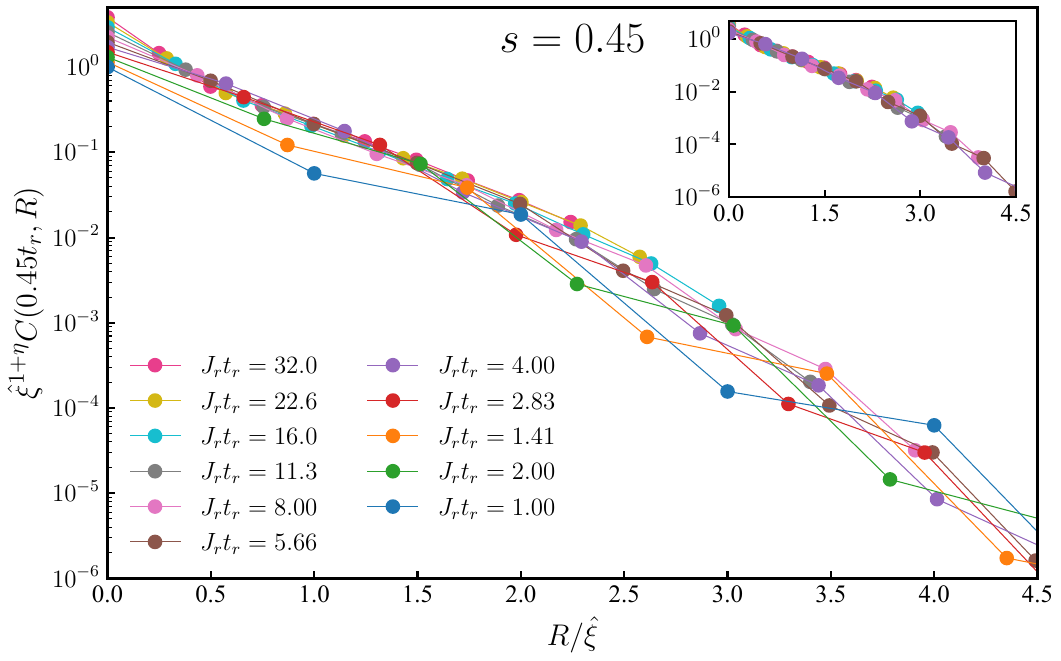}
    \includegraphics[width=0.99\columnwidth]{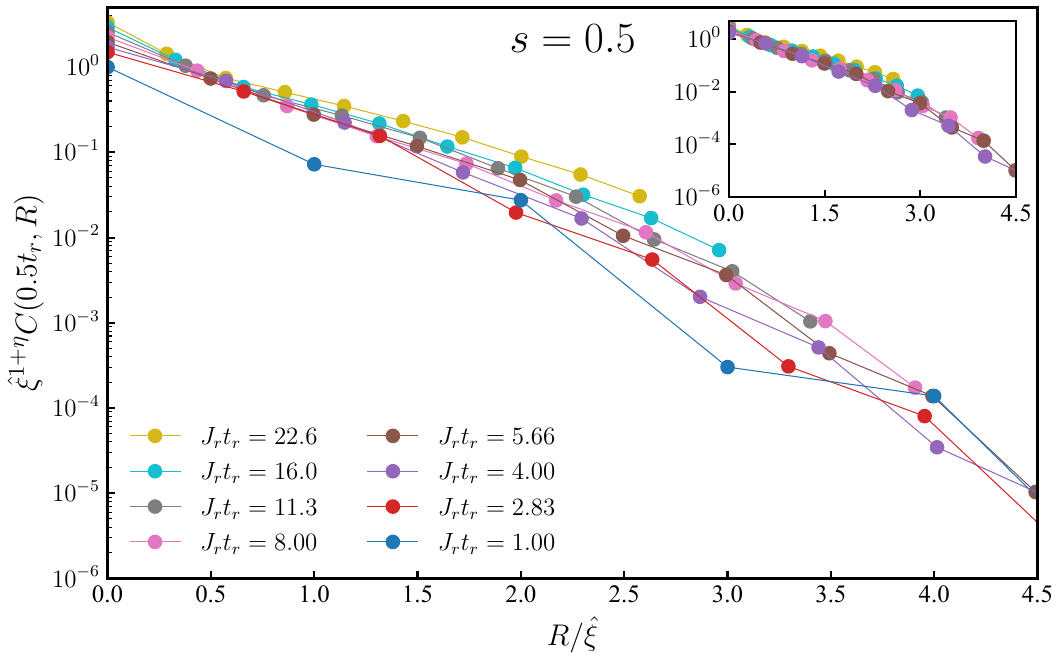}  
    \caption{
    {\bf Correlations at the critical point.}
    The scaled correlation function $\hat\xi^{1+\eta}C(t,R)$ as a function of the scaled distance $R/\hat\xi$ for several values of the ramp parameter $s=t/t_r$ near the critical point $s_c=0.45(8)$.
    At the critical $s$, with increasing ramp time $J_rt_r$, the scaled plots are expected to become smoother and collapse to a single scaling function in accordance with the Kibble-Zurek scaling hypothesis. The slower ramps ($J_rt_r\geq4$) are shown in the insets. Their collapse is the best at $s=0.45$, in consistency with the estimate for the critical point $s_c=0.45(8)$. 
    We assume this critical value, $s_c=0.45$, in the following discussion.
    }\label{fig:CR_collapse_sc}
\end{figure}

\begin{figure}[t!]
    \includegraphics[width=0.97\columnwidth]{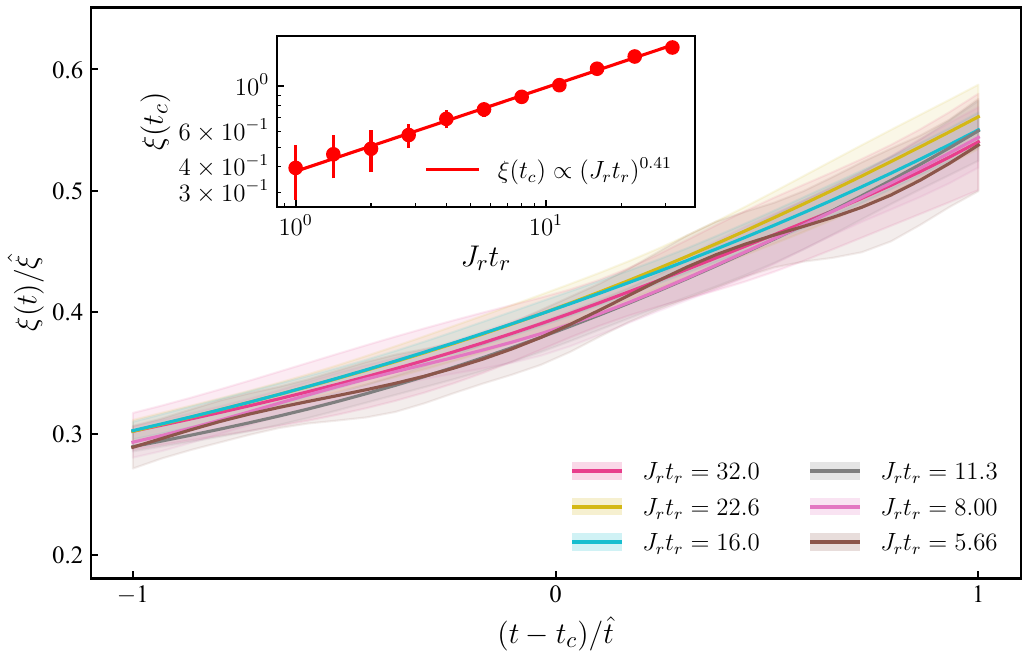}
    \caption{
    {\bf Correlation length near the critical point.}
    The scaled correlation length $\xi(t)/\hat\xi$ as a function of the scaled time $(t-t_c)/\hat t$ for different ramp times.
    Here, the KZ timescale $\hat t$ is given by $J_r\hat t = 0.36 (J_rt_r)^{z\nu/(1+z\nu)}$ with $z\nu=0.67$.
    The color-shaded areas indicate the error bars of the fit. Within the error bars, the plots collapse to a single scaling function in the KZ regime of the scaled time between $\pm 1$. 
    In particular, the inset shows the correlation length when the ramp crosses the critical point, $\xi(t_c)$, as a function of $J_rt_r$. The best fit $\xi(t_c)\propto (J_rt_r)^{0.41(3)}$ is consistent with the KZ exponent $0.40$ across a wide range of ramp times.
    }\label{fig:xi_s}
\end{figure}

\begin{figure}[t!]
    \includegraphics[width=0.97\columnwidth]{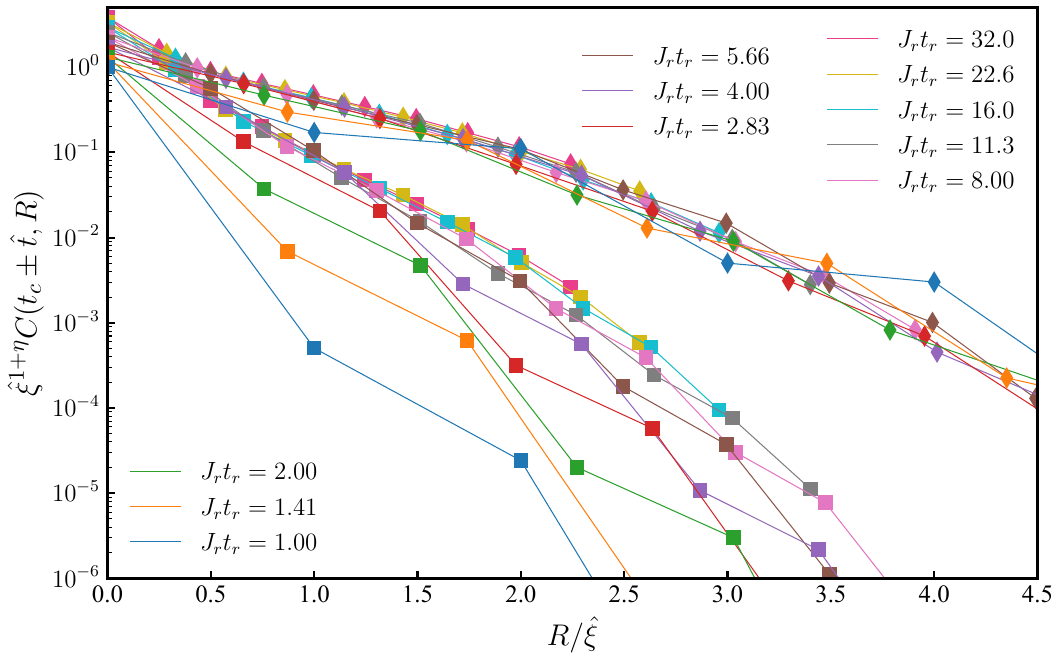}
    \caption{
    {\bf Correlations near the ends of the KZ stage.}
    The scaled correlation function $\hat\xi^{1+\eta}C(t_c\pm \hat t,R)$ in \eqref{eq:CR} at the ends of the KZ stage plotted as a function of the scaled distance $R/\hat\xi$. 
    Data marked by squares (diamonds) represent the scaled correlation functions at $t_c-\hat{t}$ ($t_c+\hat{t}$) for different ramp times.
    With increasing ramp time, the scaled correlators collapse to common scaling functions, which differ at $t_c\pm\hat t$.
    }\label{fig:CR_collapse_pm_hats}
\end{figure}

\section{Infinite lattice}
\label{sec:infty}

The central role is played by the connected staggered correlation function:
\bea 
C(t,R) 
   &=&
   \frac{(-1)^R}{2}
   \left[
   \langle \sigma^x_0 \sigma^x_R \rangle - 
   \langle \sigma^x_0 \rangle
   \langle \sigma^x_R \rangle
   \right]+          
   \left( x\to y\right).
   \label{eq:CR}
\eea
Here, the expectation value $\langle\dots\rangle \equiv \bra{\psi(t)} \dots \ket{\psi(t)}$ and $R$ is the distance along a row/column of the lattice. In order to characterize the correlation range by a single number, we fit the correlator with an exponent
\be 
C(t,R) \propto e^{-R/\xi(t)},
\label{eq:xi_fit}
\ee 
as is routinely done in many experiments. The best fit defines the correlation length $\xi(t)$. The fit is not perfect as the correlation functions are not quite exponential, and we characterize this systematic imperfection by the error bars of the fit.

The exponential fit or a fit of any other functional form, including the Ornstein-Zernike formula, is not recommended as we do not know the functional form during the diabatic part of the KZ ramp. When the correlator crosses the critical point at time $t_c$, all the scaling hypothesis \eqref{eq:KZscalingO} implies is
\be 
\hat\xi^{1+\eta} C(t_c,R) = F_C\left[0,R/\hat\xi\right]
\label{eq:CRtc}
\ee 
(compare with Fig. \ref{fig:CR_collapse_sc}) without specifying the nonuniversal scaling function $F_C$. The scaling hypothesis requires the KZ length $\hat\xi$ to be much longer than the lattice spacing $1$ and, indeed, the collapse of the data in the figure is improving with increasing ramp time. The best collapse is obtained when we assume the critical point at $s_c=0.45$, although it is not much worse at $s_c=0.45\pm0.05$. In principle, the KZ collapse could provide a more accurate estimate of the critical point if numerical simulations were able to achieve longer ramp times. For the available ramp times, the precision is comparable to the $s_c=0.45(8)$ estimated from the ground states. From now on, we fix $s_c=0.45$ for definiteness.

With all the reservations, we can still play with the fitted correlation length, as is done in Ref. \onlinecite{XX_Google}. The inset in Fig. \ref{fig:xi_s} shows the fitted $\xi(t_c)$ at the time corresponding to $s_c=0.45$. The power law is $\xi(t_c)\propto (J_rt_r)^{0.41(3)}$, in consistency with the predicted exponent $0.40$.
In the same vein, the general scaling hypothesis \eqref{eq:KZscalingO} implies further that the length should satisfy
\be 
\frac{\xi(t)}{\hat\xi} = f_\xi\left[\frac{t-t_c}{\hat\xi^z}\right].
\label{eq:fxi}
\ee 
Here, $f_\xi$ is a nonuniversal scaling function. Figure \ref{fig:xi_s} supports this hypothesis in the KZ regime within the fitting error bars. 

The KZ stage can also be probed without any reference to the fitted correlation length. In particular, the scaling hypothesis \eqref{eq:KZscalingO} implies near the ends of the KZ stage,
\be 
\hat\xi^{1+\eta} C(t_c\pm\hat t,R) = F_C\left(\pm1,R/\hat\xi\right).
\label{eq:CRhatt}
\ee 
Here, $F_C(\pm1,x)$ are nonuniversal scaling functions of the scaled distance $x$. We test this prediction in Fig. \ref{fig:CR_collapse_pm_hats} and find that, indeed, the collapse improves with increasing ramp time and increasing $\hat\xi$.

\begin{figure}[t!]
    \includegraphics[width=0.99\columnwidth]{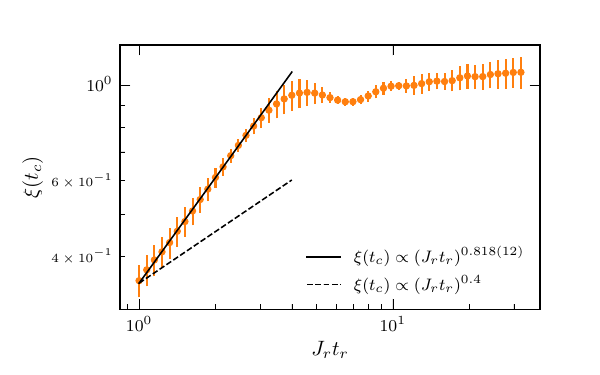}
    \caption{
    {\bf Correlation length on an $\mathbf{8\times8}$ lattice. }
    The correlation length at the critical point $\xi(t_c)$ as a function of the ramp time $J_rt_r$.
    The apparent power law for fast ramps has twice the predicted KZ exponent $0.4$.
    }\label{fig:mps_6}
\end{figure}

\begin{figure}[t!]
    \includegraphics[width=0.99\columnwidth]{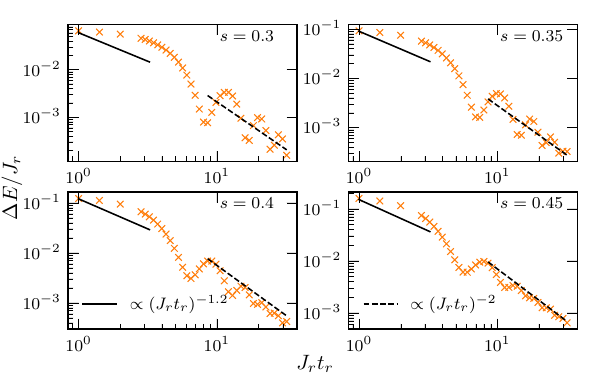}
    \caption{
    {\bf Excitation energy on an $\mathbf{8\times8}$ lattice. } 
    The total excitation energy divided by the number of sites for different ramp parameters $s$. Here, $\Delta E(s) = E(s) - E_\mathrm{GS}(s)$, where $E_\mathrm{GS}$ is the ground state energy obtained from density matrix renormalization group (DMRG) simulation of an $8\times 8$ lattice with $D=512$ for the ramp parameter $s$.
    The KZ excitation energy at the critical $s_c=0.45$ should scale as $\Delta E(s_c) \propto (J_rt_r)^{-1.20}$. 
    The discontinuous time derivative of the ramp results in $\Delta E(s) \propto (J_rt_r)^{-2}$.
    }\label{fig:mps_4}
\end{figure}

\section{$\mathbf{8\times 8}$ lattice}
\label{sec:finite}

By the collapse of the correlation functions according to the KZ scaling hypothesis, the infinite-lattice simulations demonstrate that it is possible to obtain the nonadiabatic KZ power laws in the considered range of ramp times. In this section, we use TDVP to simulate the ramp on a finite $8\times8$ square lattice with open boundary conditions. This setup is similar to the experimental one with 69 qubits. It is shown that finite-size effects result in a crossover to the adiabatic regime already for relatively fast ramps.

Indeed, Fig. \ref{fig:mps_6} shows the correlation length $\xi(t_c)$ extracted from exponential fits to the correlation function. Initially, the length increases with the ramp time, but beyond $J_rt_r\approx4$, it approximately saturates at $\xi(t_c)\approx1$. The apparent power law for faster ramps has twice the predicted KZ exponent. This result is at odds with the experimental data in Fig. 3(h) in Ref. \onlinecite{XX_Google}, where the length appears to satisfy a KZ-like power law.

In order to corroborate the adiabatic crossover, Fig. \ref{fig:mps_4} shows the excitation energy during the ramp for several ramp parameters $s$, including the critical $s_c=0.45$. 
For the dynamical exponent $z=1$, the KZ excitation energy at the critical point is expected to scale as
\be 
\Delta E \propto \hat\xi^{-(d+z)} \propto (J_rt_r)^{-1.2}.
\ee 
In contrast, when the excitation originates from a discontinuous time derivative of the ramp in a gapped system, then its energy should scale as $\Delta E\propto(J_rt_r)^{-2}$. For a simple derivation within the adiabatic perturbation theory, see the footnote on p. 20 of Ref. \onlinecite{d2010-a}. 

As the gap at $s_c$ is finite due to the finite size of the lattice, the evolution must become adiabatic for slow enough ramps. This does not mean that the excitation energy should decay exponentially with the ramp time, as might have been expected. The ramp is smooth, but the measurement of the excitation energy in the adiabatic basis of the Hamiltonian is formally equivalent to abruptly stopping the ramp. When the ramp is stopped at a finite adiabatic gap, the power-law energy decay with the exponent $-2$ is expected. 
In our simulations, the crossover to this adiabatic decay begins above $J_rt_r\approx 4$, coinciding with the saturation of the correlation length in Fig. \ref{fig:mps_6}. A similar crossover was obtained in the finite-lattice simulations of Ref. \onlinecite{schmitt2021quantum}.

The adiabatic crossover is not quite unexpected for the limited $8\times 8$ system, but it is at odds with the experimental data in Figs. 3(h) and 3(i) in Ref. \onlinecite{XX_Google}, where they seem to follow nonadiabatic KZ-like power laws. Our numerical results cannot be simply fitted with any power laws because, within the ramp times considered, there is a clear crossover to the adiabatic regime, where the correlation length saturates and the excitation energy follows the adiabatic power law.

\section{Conclusion}
\label{sec:conclusion}

Our tensor network simulations of an infinite lattice demonstrate the predicted KZ power law for correlations across a wide range of ramp times. Similar simulations of a finite $8\times8$ lattice reveal strong finite-size effects that affect the Kibble-Zurek mechanism near the quantum critical point. The slower ramps cross over into the adiabatic regime. 
The crossover is at odds with the experimental data in Ref. \onlinecite{XX_Google}, where the KZ-like power laws seem to hold.

\vspace{9pt}
The data used for the figures in this article are openly available from the RODBUK repository at Ref.~\onlinecite{UJ/ZBABII_2025}.

\acknowledgments
This research was 
funded by the National Science Centre (NCN), Poland, under Project No. 2021/03/Y/ST2/00184 within the QuantERA II Programme that has received funding from the European Union’s Horizon 2020 research and innovation programme under Grant Agreement No 101017733 (YZ),
funded by the National Science Centre (NCN), Poland, under Projects No. 2024/55/B/ST3/00626 (JD) 
%
%
and No. 2020/38/E/ST3/00150 (FB). 
The research was also supported by a grant from the Priority Research Area DigiWorld under the Strategic Programme Excellence Initiative at Jagiellonian University (JD).

\bibliographystyle{apsrev4-2}
\bibliography{KZref.bib}


\end{document}